\documentclass[aps, showpacs, showkeys,nofootinbib,floatfix]{revtex4}

\usepackage{amssymb}
\usepackage{amsmath}
\usepackage{graphicx}
\usepackage{hyperref}

\begin{document}

\title{The Numerical Solution of Scalar Field for Nariai Case in 5D Ricci-flat SdS Black String Space with Polynomial Approximation}
\author{Chunxiao Wang}
\email{chunxiaowang1231@sina.com}
\author{Molin Liu}
\email{mlliudl@student.dlut.edu.cn}
\author{Hongya Liu}
\email{hyliu@dlut.edu.cn}
\affiliation{School of Physics and
Optoelectronic Technology, Dalian University of Technology, Dalian,
116024, P. R. China}

\begin{abstract}
As one exact candidate of the higher dimensional black hole, the 5D
Ricci-flat Schwarzschild-de Sitter black string space presents
something interesting. In this paper, we give a numerical solution
to the real scalar field around the Nariai black hole by the
polynomial approximation. Unlike the previous tangent approximation,
this fitting function makes a perfect match in the leading
intermediate region and gives a good description near both the event
and the cosmological horizons. We can read from our results that the
wave is close to a harmonic one with the tortoise coordinate.
Furthermore, with the actual radial coordinate the waves pile up
almost equally near the both horizons.
\end{abstract}



\maketitle

\section{Introduction}
Our notion about the universe and the structure of the spacetime in
which we live have been radically changed by the existence of
additional spacelike dimensions in nature. It is well known that the
hierarchy problem, which is why the characteristic scale of gravity
($M_P\sim10^9GeV$) is about 16 orders of magnitude larger than the
Electro-Weak scale ($W_{EW}\sim1TeV$), is solved successfully both
in ADD model \cite{ref:ADD} and RS model \cite{ref:RS2-b}
\cite{ref:RS1-b}. In the former \cite{ref:ADD}, Arakani-Hamed et al.
suggest that compact extra dimensions can be as large as
submillimeter scale and our 4D world is a lower-dimensional brane
where all the matter is concentrated. That is, except for gravitons
and scalar particles without charges are allowed to propagate in the
bulk, all the other standard model particles are compressed on the
3-brane. Soon after, Randall and Sundrum present a different
proposal that a noncompact spacelike fifth dimension is permitted to
exist. RS 2-brane model \cite{ref:RS2-b} with a small extra
dimension and RS 1-brane model \cite{ref:RS1-b} with a large extra
dimension are build up. As a result of the instability of the vacuum
in a strong gravitational field, the black hole can radiate various
particles \cite{ref:Hawking}. People usually call it Hawking
Radiation or Black Hole Radiation. Based upon this momentous theory,
many works are explored about the entroy of black holes (see
\cite{ref:Bekenstein}-\cite{ref:SQWu}), the quasi-normal models (for
a review, see \cite{ref:Kokkotas} \cite{ref:Nollert}
\cite{ref:Jing}), and so on.

The 5D Ricci-flat Schwarzschild-de sitter (SdS) black hole, which
contains a induced cosmological constant, originates from the
canonial 1-body solution in the background of the
Space-Time-Matter (STM) theory
(\cite{ref:Wesson01}-\cite{ref:Mashhoon03}). Recently, this
solution is embedded into a brane world by using a conformial
factor \cite{ref:mlLiu01} and constitute a 5D Ricci-flat SdS black
string space. This black string space is the RS 2-brane model.
Surfaces of constant $y$, which is the fifth dimension coordinate,
satisfy the Israel equations. A reflection symmetric domain wall
is in this spacetime. The potential surrounding the black string
has a quantized spectrum as well as a continuous one by
considering the standing wave condition between its two branes.
Soon, with the tangent approximation, the real scalar field
equation is solved numerically in paper \cite{ref:mlLiu02}. Then
the extreme case--- Nariai black hole is studied in the later work
\cite{ref:mlLiu03}. The decay of scalar field is more and more
fierce with the increasing induced cosmological constant. In this
paper, we attempt to study Nariai case with a new approach
--- polynomial approximation.

Nariai solution was presented by Nariai in the 1950s
\cite{ref:Nariai} and discussed by many literatures as one of
important cases in the singular spaces \cite{ref:5articles}. It is
the exact solution to the Einstein equation with positive
cosmological constant without a Maxwell field. Meanwhile, it has a
topological structure of a (1+1)-dimensional de Sitter spacetime
with a round 2-sphere of fixed radius, i.e. $dS_2\times S^2$. Here
we study the massless scalar field about this extreme case in the
5D Ricci-flat SdS black string space. Using a polynomial as the
fitting function which is more accurate than previous tangent
approximation \cite{ref:mlLiu03}, we solve the full boundary
problem numerically.

This paper is organized as follows: in section II, we introduce the
background of this model. The 5D black string space and the two
horizons of the space are presented. In section III, assuming the
scalar field is separable, we decompose the scalar field equation
successfully. In section IV, the master propagating equation with
Schr\"{o}dinger-like form is obtained via the tortoise coordinate
transformation . In section V, with the boundary conditions and the
polynomial approximation, we obtain a full boundary value problem.
Then the Schr\"{o}dinger-like equation is solved numerically. We
finish with a summary of our result in the last section. We use the
metric signature with diagonal $(+,-,-,-,-)$ and put $\hbar$, $c$,
and $G$ equal to unity. Lower-case Greek letters $\mu$, $\nu$,
$\ldots$ will be taken to run over $0,1,2,3$ as usual, while
upper-case Latin letters $A, B, \ldots$ runs over all five
coordinates $0,1,2,3,4$.

\section{5D Ricci-flat Black String Space}
We start our analysis by presenting an exact 5D black hole solution.
The line element of the 5D Ricci-flat SdS black hole takes the form
\cite{ref:Mashhoon01}
\begin{equation}
dS^2=\frac{\Lambda
\xi^2}{3}\left[f\left(r\right)dt^2-\frac{1}{f\left(r\right)}dr^2-r^2\left(d\theta^2+\sin^2\theta
d\varphi^2\right)\right]-d\xi^2,\label{5Dmetric}
\end{equation}
where
\begin{equation}
f\left(r\right)=1-\frac{2M}{r}-\frac{\Lambda}{3}r^2,\label{f-function}
\end{equation}
 $\xi $ is the open non-compact extra dimension coordinate, $M$ is
the mass of the black hole, and $\Lambda$ is the induced
cosmological constant. The original aim of solution
(\ref{5Dmetric}), which is one of important results of STM theory,
is used to solve the canonical 1-body problem which is the 1-body
solution cannot be ruled out by solar system tests
\cite{ref:kalligas}. Considering the 5D uniqueness theorem, metric
(\ref{5Dmetric}) is introduced to describe the field outside an
object like the Sun.

This solution satisfies the 5D vacuum equation
\begin{equation}
R_{AB}=0.\label{vacuEq}
\end{equation}
Therefore, we can deduce a notable character of this space from
(\ref{5Dmetric}) and (\ref{vacuEq}). That is, the space is empty (no
matter) when viewing from the 5D. However, viewing from the 4D
(corresponds to $\xi=constant$), the matter is induced from the
empty 5D manifold. So there is no cosmological constant, but only an
effective cosmological constant $\Lambda$ induced from the fifth
dimension.

Mathematically, the three solutions of
\begin{equation}
f(r)=0\label{f=0}
\end{equation}

can be expressed \cite{ref:Liu01} as:
\begin{equation}
\left\{
\begin{array}{l}
r_c=\frac{2}{\sqrt{\Lambda}}\cos\eta,\\
r_e= \frac{2}{\sqrt{\Lambda}}\cos(120^\circ-\eta),\\
r_o=- \left( {r_e  + r_c } \right) \\
\end{array}
\right.\label{re-rc}
\end{equation}
where $\eta=\frac{1}{3}\arccos(-3M\sqrt{\Lambda})$ with $30^\circ
\leq\eta\leq 60^\circ$, and $\Lambda$ must satisfy $\Lambda
M^2\leq\frac{1}{9}$ to ensure there are acceptable solutions to
(\ref{f=0}). We have labeled the black hole horizon (event horizon)
and the cosmological horizon with $r_e$ and $r_c$ respectively.
$r_o=-(r_e+r_c)$ is a negative solution and has no actual
significance.

The space (\ref{5Dmetric}) is bounded by the two horizons. The
interval between them is decreased as increasing $\Lambda$. The
Nariai black hole arises when two horizons close to each other. All
through the paper we will consider this extreme case. Consequently,
the induced cosmological constant is chosen $\Lambda=0.11$ (where
have taken $M=1$) like the valuation in \cite{ref:Tian}
\cite{ref:Brevik}.

Expression (\ref{f-function}) can be rewritten as follows
\begin{equation}
f(r)=\frac{\Lambda}{3r}(r-r_{e})(r_{c}-r)(r-r_{o}). \label{ref-fun}
\end{equation}
Redefining the fifth dimension through the following transform
\begin{equation}
\xi=\sqrt{\frac{3}{\Lambda}}e^{\sqrt{\frac{\Lambda}{3}}y},\label{replacement}
\end{equation}
we can get the new form of line element (\ref{5Dmetric}) as
\begin{equation}
dS^{2}=e^{2\sqrt{\frac{\Lambda}{3}}y}\left[f\left(r\right)dt^{2}-\frac{1}{f\left(r\right)}dr^2-r^2\left(d\theta^2+\sin^2\theta
d\varphi^2\right)-dy^{2}\right],\label{eq:5Dmetric-y}
\end{equation}
where $y$ is the new fifth dimension coordination.
 \section{Klein-Gordon equation of massless scalar field in the bulk}
Now we turn our attention to a massless scalar filed $\phi$ in the
5D black string space. The field
$\phi\left(t,r,\theta,\varphi,y\right)$ obeys the Klein-Gordon
equation
\begin{equation}
\square\phi=0.
\end{equation}
By using the field factorization
\begin{equation}
\phi=\frac{1}{\sqrt{4\pi\omega}}\frac{1}{r}R_{\omega}(r,t)L(y)Y_{lm}(\theta,\varphi)\label{wave
function}
\end{equation}
and
\begin{equation}
R_{\omega}(r,t)\rightarrow\Psi_{\omega l n} (r) e^{-i \omega t},
 \end{equation}
where $\omega$ is the energy of scalar particles. $R_\omega(r,t)$
is the radial function about $(r,t)$, $Y_{lm}(\theta,\varphi)$ is
the usual spherical harmonic function, and $L(y)$ is the fifth
dimension wave function. We obtain a set of decoupled 5th
dimensional and radial equations
\begin{eqnarray}
 &&\frac{d^2L\left(y\right)}{dy^2}+\Lambda\sqrt{\frac{\Lambda}{3}}\frac{dL\left(y\right)}{dy}+\Omega
 L\left(y\right)=0,\label{5th-equation} \\
 &&\left[-f\left(r\right)\frac{d}{dr}\left(f\left(r\right)\frac{d}{dr}\right)+V\left(r\right)\right]\Psi_{\omega
 l n}\left(r\right)=\omega^2\Psi_{\omega l n} (r),\label{radius equ. about r}
  \end{eqnarray}
where $\Omega$, $l$ are the constants which are adopted to separate
variables. The potential function have been defined as follows:
\begin{equation}
V\left(r\right)=f\left(r\right)\left[\frac{1}{r}\frac{df\left(r\right)}
{dr}+\frac{l\left(l+1\right)}{r^2}+\Omega\right].\label{potential}
\end{equation}
According to RS 2-brane model, there are two branes with $Z_2$
symmetry in the 5D bulk. They have opposite and equal tensions, and
one of them contains the standard model fields. The two branes can
be be set at $y=0$ and $y=y_1$ (where $y_1$ is the thickness of the
bulk) respectively. Using the standing wave condition between the
two branes, we can solve Eq. (\ref{5th-equation}) and get the
spectrum of $\Omega$, which includes continuous part below
$\frac{3}{4}\Lambda$ and discrete part above $\frac{3}{4}\Lambda$
(about detail analysis, refer to \cite{ref:mlLiu01}). The quantum
parameter is denoted by $\Omega_n(n = 1, 2, 3, \ldots)$:

\begin{equation}
\Omega_n=\frac{n^2\pi^2}{y_1^2}+\frac{3}{4}\Lambda
\end{equation}

\section{tortoise coordinate and master propagating equation}
\begin{figure}
\includegraphics*[width=3.5in]{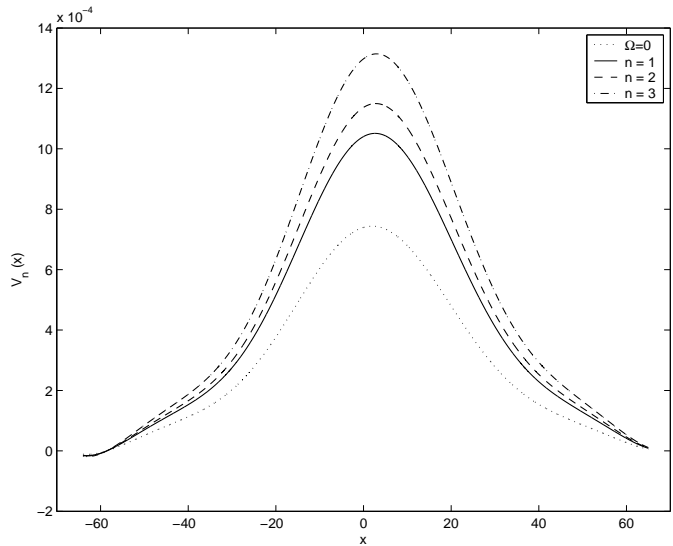}
 \caption{The first 3 ($n=1,2,3$) quantum potential barriers $V(r)$ for 5D string
   space. Here we selected $M=1,\omega=1,l=1,\Lambda=0.11$, and
   $y_{1}=10^{\frac{3}{2}}$ (a large extra dimension). The dotted line ($\Omega=0$) corresponds to the 4D SdS
   case.}\label{tu1}
\end{figure}

Here we use the tortoise coordinate, defined by the
relation
\begin{equation}
x=\frac{1}{2M}\int{\frac{dr}{f\left(r\right)}}.\label{tortorse}
\end{equation}
By using (\ref{f-function}), expression (\ref{tortorse}) can be
integrated and the tortoise coordinate can be written as
\begin{equation}
x=\frac{1}{2M}\left[\frac{1}{2K_e}\ln\left(1-\frac{r}{r_e}\right)-
\frac{1}{2K_c}\ln\left(1-\frac{r}{r_c}\right)
+\frac{1}{2K_o}\ln\left(1-\frac{r}{r_o}\right)\right],\label{anti-tot}
\end{equation}
where the surface gravities $K_i$ is defined as:
\begin{equation}
K_i=\frac{1}{2}\left|\frac{df}{dr}\right|_{r=r_i}.
\end{equation}
So the master propagating Eq. (\ref{radius equ. about r}) takes the
Schr$\ddot{o}$dinger-like form
\begin{equation}
\left[-\frac{d^2}{dx^2}+4M^2V\left(r\right)\right]\Psi_{\omega n
l}\left(r\right)=4M^2\omega^2\Psi_{\omega n
l}\left(r\right),\label{SchEq}
\end{equation}
and the gravitational potential barrier $V(r)$ that a scalar
particles sees while propagating from the cosmological to the black
hole horizon or vice versa takes the form
\begin{equation}
V\left(r\right)=f\left(r\right)\left[\frac{1}{r}\frac{df\left(r\right)}
{dr}+\frac{l\left(l+1\right)}{r^2}+\frac{n^2\pi^2}{y_1^2}+\frac{3}{4}\Lambda\right].\label{QutPoten}
\end{equation}
As an illuminating example, Fig. \ref{tu1} depicts the form of this
barrier for $n=1, 2, 3$.
\section{A Full Boundary Value Problem}
\subsection{Boundary Conditions and Polynomial Approximation}
In order to solve the wave equation (\ref{SchEq}) numerically, we
need the boundary conditions at the horizons. Since the metric
function $f (r)$ vanishes at both horizons, $r_e$ and $r_c$, so does
the potential barrier that is proportional to the metric function.
Therefore, in both of these regimes, the solution of Eq.
(\ref{SchEq}) has the form of plane waves, that is $\\e^{\pm
i2M\omega x}$. Being consider only real field there, we can choose
the boundary conditions at both horizons as
\begin{equation}
\Psi_{\omega l n}\left(x\right)=\cos\left({2M\omega
x}\right).\label{bdary1}
\end{equation}
Before solve the Eq. (\ref{SchEq}), we need the potential is a
function of of tortoise coordinate, i. e. $V=V(x)$. But it is
difficult to invert Eq. (\ref{anti-tot}) to get $r=r(x)$. So we can
adopt the polynomial approximation
\begin{equation}
y=\tilde{r}=\sum\limits_{i=0}^{10}{a_ix^i}\label{fitting}
\end{equation}

\begin{table}[!b]
\caption{The coefficients in the approximating polynomial of degree
10}
\begin{tabular}{|l|l|l|}
     \hline\hline
     $a_0=2.9821$&$a_1=6.4016\times10^{-3}$&$a_2=3.4629\times10^{-5}$\\
     \hline
     $a_3=-2.3878\times10^{-6}$ &$a_4=-1.7683\times10^{-8}$&$a_5=6.5626\times10^{-10}$\\
     \hline
     $a_6=4.5015\times10^{-12}$ &$a_7=-8.9336\times10^{-14}$  &$a_8=-5.5464\times10^{-16}$\\
     \hline
    $a_9=4.3675\times10^{-18}$ &$a_{10}=2.7548\times10^{-20}$&\\
     \hline
     \hline
\end{tabular}\label{table}
\end{table}
to approximate tortoise coordinate. The coefficients ${a_i}$ are
showed in Table \ref{table}.

\begin{figure}
\includegraphics*[width=3.5in]{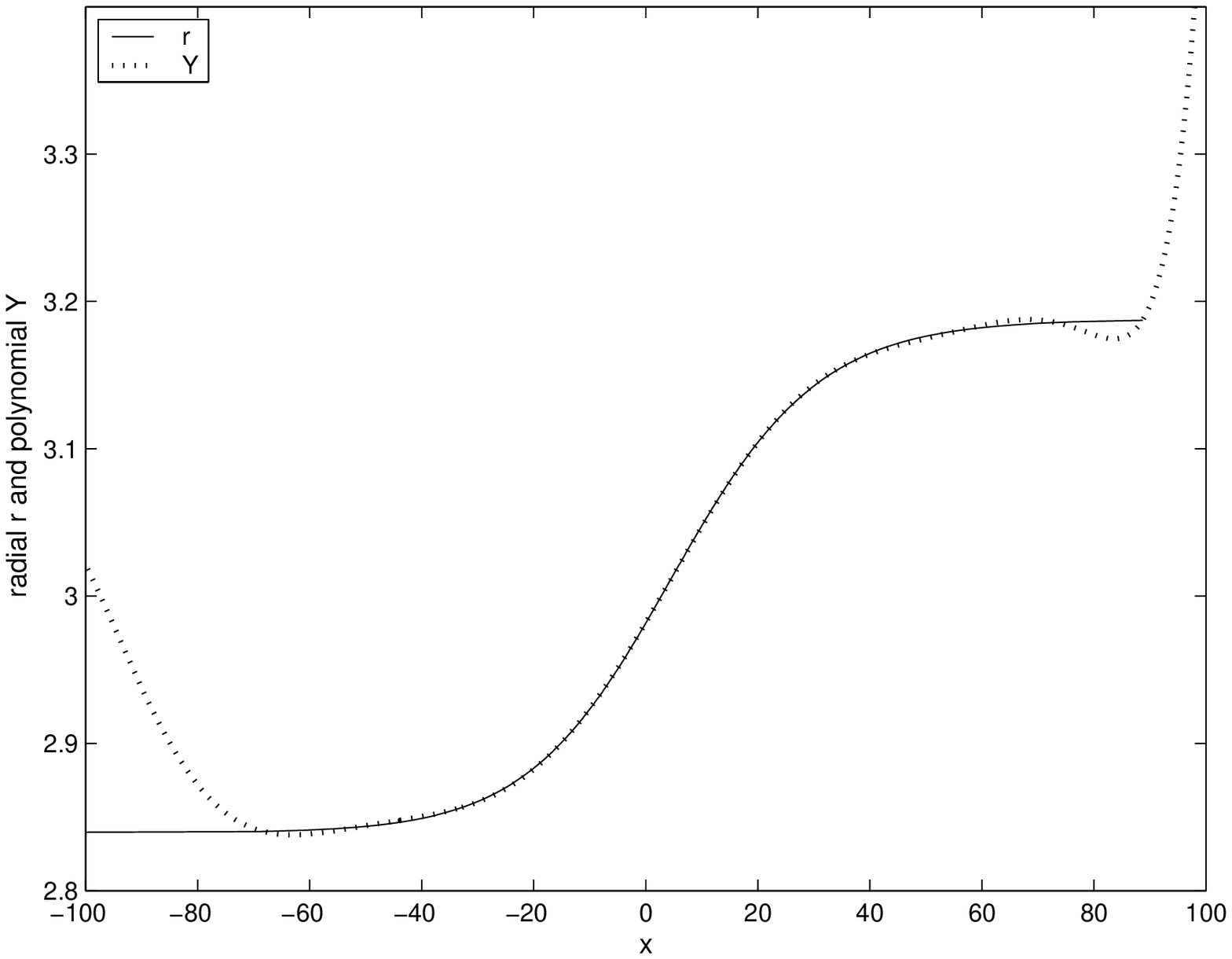}
 \caption{Radial coordinate $r$ (solid) and its polynomial approximate $y$
 (dotted)} versus the tortoise coordinate $x$ with $\Lambda=0.11$.\label{tu2}
\end{figure}

As an illustration of the accuracy of the approximation, we plot the
curves of $r$ and $y$ (or $\tilde{r})$ versus $x$ in Fig. \ref{tu2},
from which we notice that approximation (\ref{fitting}) does not
allow $\left|x\right|$ to become very large, so we shorten the
interval of $x$ to $\left[-70,70\right]$, and the boundary condition
(\ref{bdary1}) is rewritten as
\begin{equation}
\Psi_{\omega l n}\left(-70\right)=\Psi_{\omega l
n}\left(70\right)=\cos\left({140M\omega}\right).\label{bdary2}
\end{equation}
It is to be noted here that the interval $[-70,\ 70]$ is selected
according to the fitting situation. This setting can be
illustrated clearly in Fig. \ref{tu2}. As far as we know, there is
no way to give the accurate description near horizons in SdS black
hole whatever in 4D or higher dimensional space.  Comparing with
the tangent approximation \cite{ref:Brevik} \cite{ref:mlLiu02}
\cite{ref:mlLiu03}, the polynomial approximation shows good
representing in despite of the shorter fitting interval $[-70,\
70]$ than the origin one $[-100,\ 100]$ \cite{ref:Brevik}.

\subsection{Numerical Solution of the Master Propagating Equation}
\begin{figure}
\includegraphics*[width=3.5in]{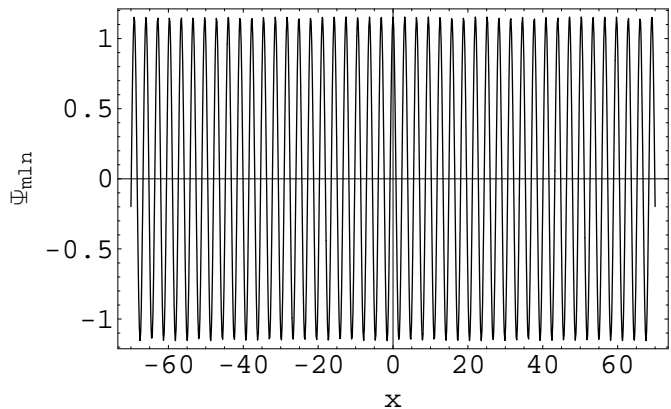}
 \caption{Variation of the field amplitude versus $x$ with $M=1,\omega=1,l=1,\Lambda=0.11,
 y_1=10^{\frac{3}{2}}$ and $n=1$. The solution is close to a harmonic wave.}\label{tu3}
\end{figure}

\begin{figure}
\includegraphics*[width=3.5in]{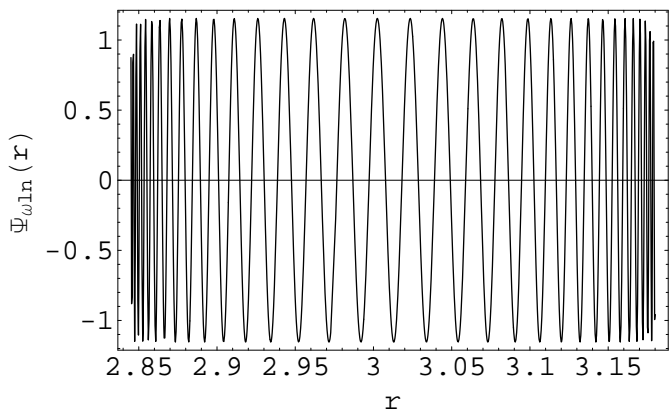}
 \caption{Variation of the field amplitude versus $r$ with $M=1,\omega=1,l=1,\Lambda=0.11,
 y_1=10^{\frac{3}{2}}$ and $n=1$. The waves pile up equally near both horizons.}\label{tu4}
\end{figure}

The radial wave equation (\ref{SchEq}) and the boundary condition
(\ref{bdary2}) can build up a full boundary value problem.
Considering the polynomial approximation (\ref{fitting}), we can
solve the problem numerically by Mathmatica software. The solution
for the first quantum state ($n=1$) is illustrated in Fig. \ref{tu3}
and Fig. \ref{tu4}. And the other states can be obtained by the same
way.

Fig. \ref{tu3} illustrates the behavior of the field amplitude
between the horizons when $x$ is the independent variable. It is
shown that the solution is close to a harmonic wave. With $r$ as
independent variable, Fig. \ref{tu4} shows that the waves pile up
and the wavelength goes to zero near the two horizons.

Because of the compact property of the tortoise coordinate, it is
natural that the waves pile up near the both horizons. The different
behavior of the waves  near the two horizons in \cite{ref:mlLiu03}
is caused mainly by the imprecise tangent approximate method. Since
the same approximate method is used in \cite{ref:Brevik}, we can
compare our Fig. 2 directly with the Fig. 3 of \cite{ref:Brevik} and
conclude that our fitting effect is more precise. Our result shows
that the densities of the wave near the two horizons of Fig.
\ref{tu4} are nearly equal in both horizons.

\section{Conclusion}
The 5D Ricci-flat SdS black string space is derived from a 5D Ricci
flat manifold in the STM scenario. The construction of this space is
that two 3-branes are embedded into a empty bulk like RS 2-brane
model \cite{ref:RS2-b}. But unlike the slab of Anti-de Sitter (AdS)
space in the standard Randall-Sundrum  two branes system, there is
no cosmological constant but only the induced (effective) one in
this 5D bulk. It is worthy mentioning that the quantum potential
labeled by the quantum number $n$ is one of the important results in
this black string space. The height and the thickness of the
potential are bigger with the increasing quantum number. We assume
that the massless scalar field without charge can propagate free in
the bulk, and the two branes are stabilized by this scalar field.
Furthermore, we consider the Nariai black hole, which is one of the
important cases in the singular spaces. In this paper, we employ a
more accurate fitting function--- polynomial approximation to fit
the radial coordinate. With this kind of approximation, a full
boundary value problem is formed and solved successfully. Because of
the complexity of the differential equation (\ref{SchEq}), only the
numerical solution is presented here. Comparing with the previous
results in \cite{ref:mlLiu03}, we can find that the polynomial
approximation is more reasonable and the waves pile up almost
equally near both horizons.

\acknowledgments
This work was supported by NSF (10573003) and NBRP
(2003CB716300) of P. R. China.

\end{document}